\documentclass[12pt]{article}
\usepackage{amsmath}
\usepackage{mitpress}

\newdimen\dummy
\dummy=\oddsidemargin
\input{tcilatex}

\begin{document}

\title{Blackbody radiation drag on a relativistically moving mirror}
\author{\textbf{N.R.Balasanyan and V.E.Mkrtchian} \\
Institute for Physical Research, Armenian Academy of Sciences, Ashtarak-2,
0410, Republic of Armenia.}
\maketitle

\begin{abstract}
We compute the drag force on a mirror moving at relativistic velocity
relative to blackbody radiation background.

\textbf{Key words \ }Van der Waals friction, blackbody radiation,
relativistic mirror.

\end{abstract}

Recently a great deal of attention has been devoted to the problem of
frictional force acting on a neutral particle\ when it moves through
blackbody radiation \cite{vem}, \cite{Lukas}, \cite{Vol}, \cite{DK}.
However, despite of all efforts an unambiguous generalization of the results 
\cite{vem}, \cite{Lukas} in the case of relativistic velocities of the
particle at the moment \ is missing (see \cite{DK}) .

In this letter we solve blackbody radiation drag problem exactly in the case
of "structureless" moving body, i.e. for a mirror.

Let us consider a mirror moving with velocity $\mathbf{v}$ through a
thermalized photonic gas with a temperature $T.$ In the frame of the mirror
distribution function of photons is given by the expression \cite{Light}%
\begin{equation}
n\left( \mathbf{k,v}\right) =\frac{1}{e^{\gamma \left( \omega +\mathbf{kv}%
\right) /T}-1},  \tag{1}
\end{equation}%
where 
\begin{equation*}
\gamma =1/\sqrt{1-\beta ^{2}},\text{ }\beta =v/c.
\end{equation*}%
By using distribution function (1) we find the momentum density $\mathbf{p}$
of the electromagnetic field:%
\begin{equation}
\mathbf{p=}\dint \frac{2d\mathbf{k}}{\left( 2\pi \right) ^{3}}\mathbf{k}%
n\left( \mathbf{k,v}\right) =-\mathbf{v}\frac{16}{3c^{3}}\sigma T^{4}\frac{1%
}{1-\beta ^{2}},  \tag{2}
\end{equation}%
where $\sigma $ is the Stefan-Boltzmann constant 
\begin{equation*}
\sigma =\frac{\pi ^{2}k_{B}^{4}}{60\hbar ^{3}c^{2}}
\end{equation*}

Then, for a plane mirror moving perpendicularly to its surface we get for
the drag force $f$ per unite area of surface\ the following expression%
\begin{equation}
f=\frac{32}{3c}\sigma T^{4}\frac{\beta }{1-\beta ^{2}}.  \tag{3}
\end{equation}

This is the main result result of this communication and it is correct for
arbitrary large velocities of the mirror. As seen, the drag force (3) is
proportional to $T^{4}$ as in case of a nonrelativistically moving metallic
particle \cite{vem}.

It is useful to compare the drag force pressure (3) with the of blackbody
radiation pressure $P$ \cite{V}%
\begin{equation}
P=\frac{4\sigma }{3c}T^{4}.  \tag{4}
\end{equation}%
From expressions (3) and (4) we find the ratio%
\begin{equation}
f/P=\frac{8\beta }{1-\beta ^{2}}.  \tag{5}
\end{equation}

Thus, in the nonrelativistic limit of velocities $\left( \beta \ll 1\right) $
the drag force is proportional to the velocity of mirror: 
\begin{equation}
f/P\sim \beta  \tag{6}
\end{equation}%
and is very small.

For the velocities$\allowbreak $ $v\simeq 0.1c$ the drag pressure becomes
comparable with the blackbody radiation pressure:%
\begin{equation}
f/P\sim 1.  \tag{7}
\end{equation}

In the limit of relativistic velocities $\left( \beta \rightarrow 1\right) $
of the mirror the drag force density $f$ becomes infinitely large:%
\begin{equation}
f/P\sim (1-\beta )^{-1}.  \tag{8}
\end{equation}

\end{document}